\def\K{$^{40}$K }
\def\Na{$^{23}$Na}
\def\NaK{\Na\K}

\def\Xstate{X$^1\Sigma^+$}
\def\astate{a$^3\Sigma^+$}

\def\braket#1{\mathinner{\langle{#1}\rangle}}

\def\Ket#1{\left|#1\right\rangle}
\documentclass[aps,pra,twocolumn,floatfix,10pt]{revtex4} 
\usepackage{graphicx,amsmath,amssymb,ulem}
\usepackage{color}
\usepackage{nicefrac}

\begin{document}

\title{Multi-channel modeling and two photon coherent transfer paths in NaK}

\author{T.A.~Schulze$^1$}\author{I.I.~Temelkov$^{1,2}$}\author{M.W.~Gempel$^1$}\author{T.~Hartmann$^1$}\author{H.~Kn\"ockel$^1$}\author{S.~Ospelkaus$^1$}\author{E.~Tiemann$^1$}

\affiliation{$^1$ Institut f\"ur Quantenoptik, Leibniz Universit\"at Hannover, 30167~Hannover, Germany}
\affiliation{$^2$ Department of Physics, Sofia University, 5 James Bourchier Boulevard, 1164 Sofia, Bulgaria}

\date{\today}

\begin{abstract}
We explore possible pathways for the creation of ultracold polar $\mathrm{NaK}$ molecules in their absolute electronic and rovibrational ground state starting from ultracold Feshbach molecules. In particular, we present a multi-channel analysis of the electronic ground and K(4p)+Na(3s) excited state manifold of NaK, analyze the spin character of both the Feshbach molecular state and the electronically excited intermediate states and discuss possible coherent two-photon transfer paths from Feshbach molecules to rovibronic ground state molecules.  The theoretical study is complemented by the demonstration of STIRAP transfer from  the \Xstate (v=0) state to the \astate~manifold on a molecular beam experiment. 

\end{abstract}

\maketitle
\section{Introduction} \label{sec1}
Atomic Bose condensates and degenerate Fermi gases are nowadays established experimental tools for probing and simulating quantum many-body phenomena which are difficult to study in their original context. In most of the atomic systems, the interparticle interaction can be described by a contact pseudopotential. This restricts the range of possible systems which can be quantum simulated to those with short-range interaction. However, intriguing phenomena are expected to occur in systems with dipolar interaction and this leads to rapidly growing interest in the preparation and study of dipolar quantum gases. Pioneering experiments have made use of magnetic dipolar interactions of atoms with large magnetic moments such as Cr~\cite{Chromium} and more recently Er and Dy~\cite{Erbium,Dysprosium}. Even larger dipolar interactions can be realized making use of electric dipole-dipole interaction. Therefore,  recent activities have focused on the creation of ultra-cold polar molecules in their rovibronic ground state, in which dipole moments on the order of several Debye can be induced by electric fields.
\begin{figure}
\centering
\includegraphics*[width=0.9\columnwidth]{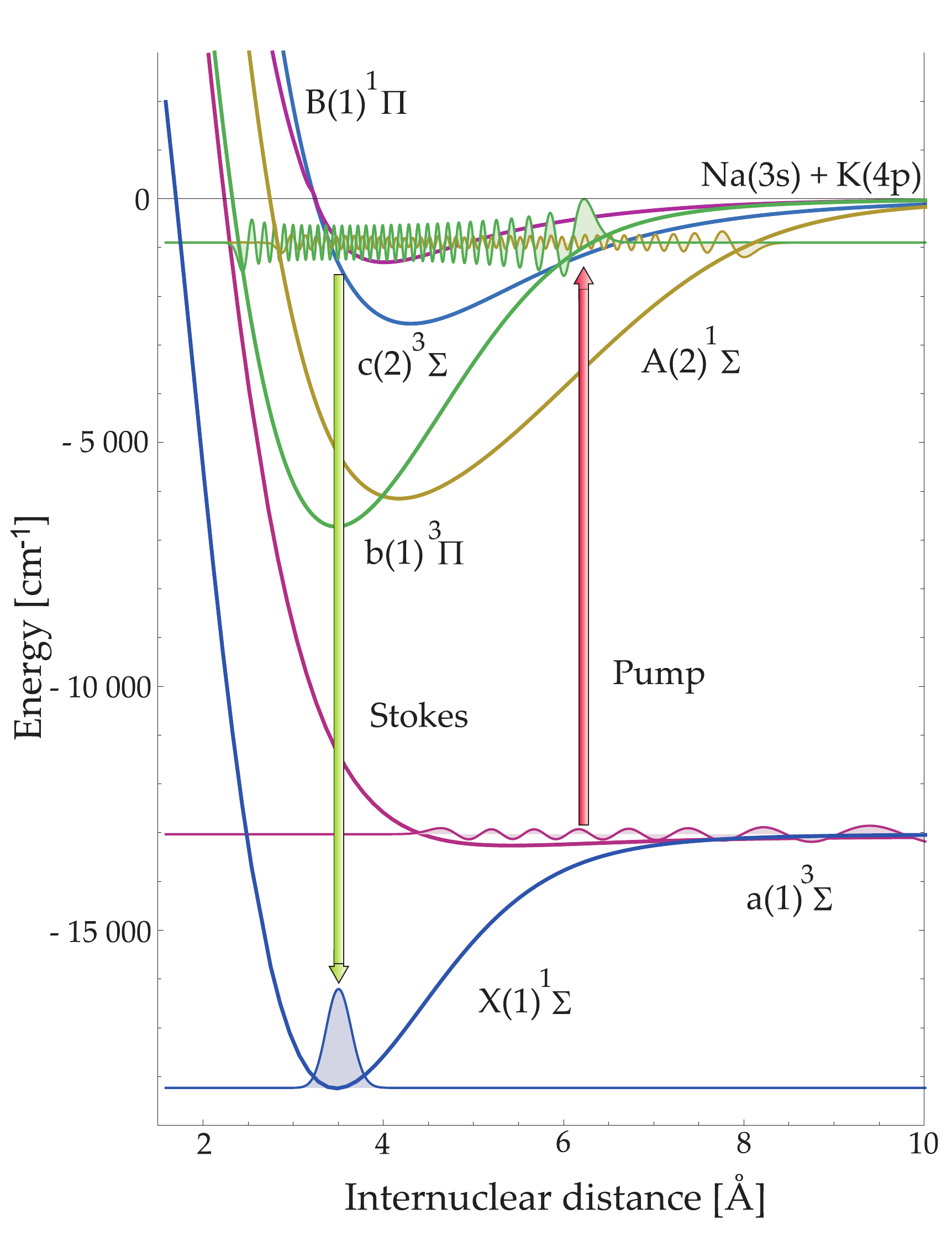}
\caption{Scheme of the coherent transfer sequence from a Feshbach level to the absolute ground state of the molecule. In this example, the intermediate level is of the resonant $A^{1}\Sigma \sim b^{3}\Pi$ type. The wavefunction amplitudes reflect the singlet or triplet components.}
\label{fig0}
\end{figure}

Due to the difficulties in directly cooling molecules, currently, the only experimental pathway for the preparation of quantum gases of molecules starts from the preparation of quantum degenerate gases of atoms followed by an association of atom pairs in the quantum gas into a rovibronic ground state molecule. In these experiments, molecule creation is being performed in a fully coherent two step process: First, atoms in the quantum gas are being associated into weakly bound Feshbach molecules in the vicinity of a Feshbach resonance. This association process is followed by a coherent two-photon transfer of these molecules into rovibronic ground state molecules. Due to the fully coherent nature of the process, the phase space density of the initial atomic ensembles is being preserved resulting in a dense ultracold molecular sample close to quantum degeneracy. 

This experimental approach for ultracold molecule creation has first been demonstrated in a seminal experiment of the JILA group~\cite{Ospel2008,Faraday}, where an ultracold dense gas of KRb molecules close to quantum degeneracy has been prepared. However, the specific alkali dimer choice in the JILA experiment invokes a severe loss channel due to exothermic chemical reactions converting two colliding KRb molecules into the corresponding homonuclear dimers. Binding energy differences being of the order of a few $\mathrm{cm}^{-1}$ translate into thermal energies of several Kelvin, this leads to massive heating, limiting the lifetime of the sample~\cite{Ospel2010,chemicalreactionsBohn}. Such a loss channel can be suppressed by choosing one of the constituent atoms of the molecules to be Cesium or Sodium, and chemical reactions as mentioned above will become endothermic~\cite{Inelastic2010}.

NaK is a promising candidate, due to the possibility to explore bosonic as well as fermionic molecules. Furthermore, it has been extensively studied by molecular spectroscopy in the past. A large number of experimental data and spectroscopic constants are available in the literature, leading e.g. to the potential energy curves (PECs) of the involved electronic states ~\cite{Russier,Zemke,Stolyarov,Kato,Barrow}, and are supplemented by sophisticated theoretical studies of molecular properties like radiative lifetimes, dipole moments or static polarizabilities~\cite{Tamanis,static}. Recently, Feshbach resonances have been observed~\cite{Zwierlein1} and ultracold fermionic \NaK Feshbach molecules created subsequently~\cite{Zwierlein2}. Yet the two-photon transfer of Feshbach molecules into rovibronic ground state molecules remains to be explored. 

Here, we present a detailed analysis of possible two-photon pathways for preparing ultracold NaK in its absolute ground state. Fig.\,\ref{fig0} shows for the ground asymptote K($4s$)+Na($3s$) and the excited asymptote K($4p$)+Na($3s$) the molecular potentials, which are involved in the envisaged two-photon transfer, and a sketch of a typical two-photon scheme for mapping Feshbach molecules onto the rovibrational ground state. The pump light field couples Feshbach molecules to an appropriate electronically excited intermediate state;  the overlapping Stokes pulse couples the target \Xstate rovibrational ground state to the same excited state level - resulting in a two-photon Raman transfer. By this type of transfer, we overcome two problems, firstly, the direct wave function overlap between weakly bound Feshbach molecules and deeply bound ground state molecules is vanishingly small, and  secondly, Feshbach molecules and ground state molecules of alkali dimers have very different electronic spin character. Whereas Feshbach molecules are often dominantly \astate~in character, rovibronic ground state molecules are purely \Xstate~molecules. Therefore, an appropriate electronically excited state serves as a bridge for the transfer - both in terms of wave function overlap and electron spin triplet-singlet mixing for triplet-to-singlet conversion. We note two differences between the KRb experiment \cite{Ospel2008} and the NaK case.  First, by comparing atomic spin-orbit coupling constant $\xi_{Rb,(K),[Na]} = 79.20 (19.24) [5.73] \mathrm{cm}^{-1}$ of atoms Rb, K and Na in their lowest p- state, one sees that such a bridge will be expected to be far weaker for NaK than KRb. As one cannot endlessly compensate this weakness by increasing laser power and focusing of light beams, the suitability of such a scheme has to be reevaluated for NaK. The second difference between NaK and KRb is hidden in the Feshbach molecular state. For NaK, broad resonances have been reported~\cite{Zwierlein1} which can vastly influence the spin character and it was stated~\cite{Zwierlein2} that the singlet amount in the Feshbach state might allow for direct singlet coupling to an intermediate state, which would render unnecessary a triplet-singlet bridge, asking also for reevaluation for the case NaK. 
Ultimately, the parameter of interest is the product of the two dipole matrix elements, representing the two-photon transition:

\begin{equation} d_{(\mathrm{F}\rightarrow \mathrm{X})} = \langle\psi_{\mathrm{F}} | \hat{d} | \psi_{\text{int.}}\rangle \langle\psi_{\mathrm{int.}} | \hat{d} | \psi_{\mathrm{X}}\rangle \label{eq:dip} \end{equation}
Here, $\Ket{\psi_{\mathrm{F}}}$ corresponds to the Feshbach molecule, whereas $\Ket{\psi_{\mathrm{int.}}}$ is the intermediate level selected for an effective two-photon process. The rovibronic ground state is abbreviated as $ \Ket{\psi_{X}} \equiv\Ket{X^{1} \Sigma^{+} (v = 0, J = 0)}$. 
This work focuses on a detailed analysis of the involved molecular potentials to identify windows where both the transition amplitude from the Feshbach state to the intermediate state and the transition amplitude from the intermediate state to the ground state are sizable. For the Feshbach state near a resonance, we show the influence of the hyperfine interaction on the singlet admixing and discuss the suitability of a pure singlet transition window. The intermediate state is studied with particular emphasis on its spin mixing characteristics. Combining the ground and excited state analysis two-photon transition dipole matrix elements are calculated and possible two-photon transfer paths discussed. Furthermore, we demonstrate population transfer between the  \Xstate~vibrational ground state and the \astate~manifold via a STIRAP sequence in a molecular beam experiment.

\section{Ground state modeling}
\label{groundstate}

We start by investigating the initial level of the two-photon sequence. This already gives insight into the requirements concerning the choice of an intermediate level in order to accomplish an efficient ground state conversion. The crunchpoint of eq. \eqref{eq:dip} lies in the radial variation of the vibrational wavefunctions, which are obtained from the corresponding PECs. Hence, detailed knowledge of the PECs is essential for our task. 
The singlet and triplet ground states of the NaK molecule were studied in earlier experiments of our group~\cite{Gerdes2008,Gerdes2011}. In the following we use a slightly updated version of the PECs compared to~\cite{Gerdes2008} for our analysis, where existing data  from Fourier transform spectroscopy have been complemented by recent results of Feshbach spectroscopy \cite{Zwierlein1}. This allows us to obtain an accurate description of the least bound states and serves our purpose in this letter very well for various reasons. First, we are able to give precise statements regarding the asymptotic part of the multicomponent wavefunction, including Feshbach resonances and molecules. Second, the joint description of the PECs fixes their relative position and therefore serves as a common frequency reference connecting the singlet and triplet manifold, removing additional uncertainties in the two-photon detuning of Pump and Stokes lasers. Last, we employ the ground state PECs as a benchmark in order to improve upon \textit{ab initio} calculations for the excited states (see later section \ref{intermediate} for details). 
\begin{figure}
\centering
\includegraphics*[width=1.0\columnwidth]{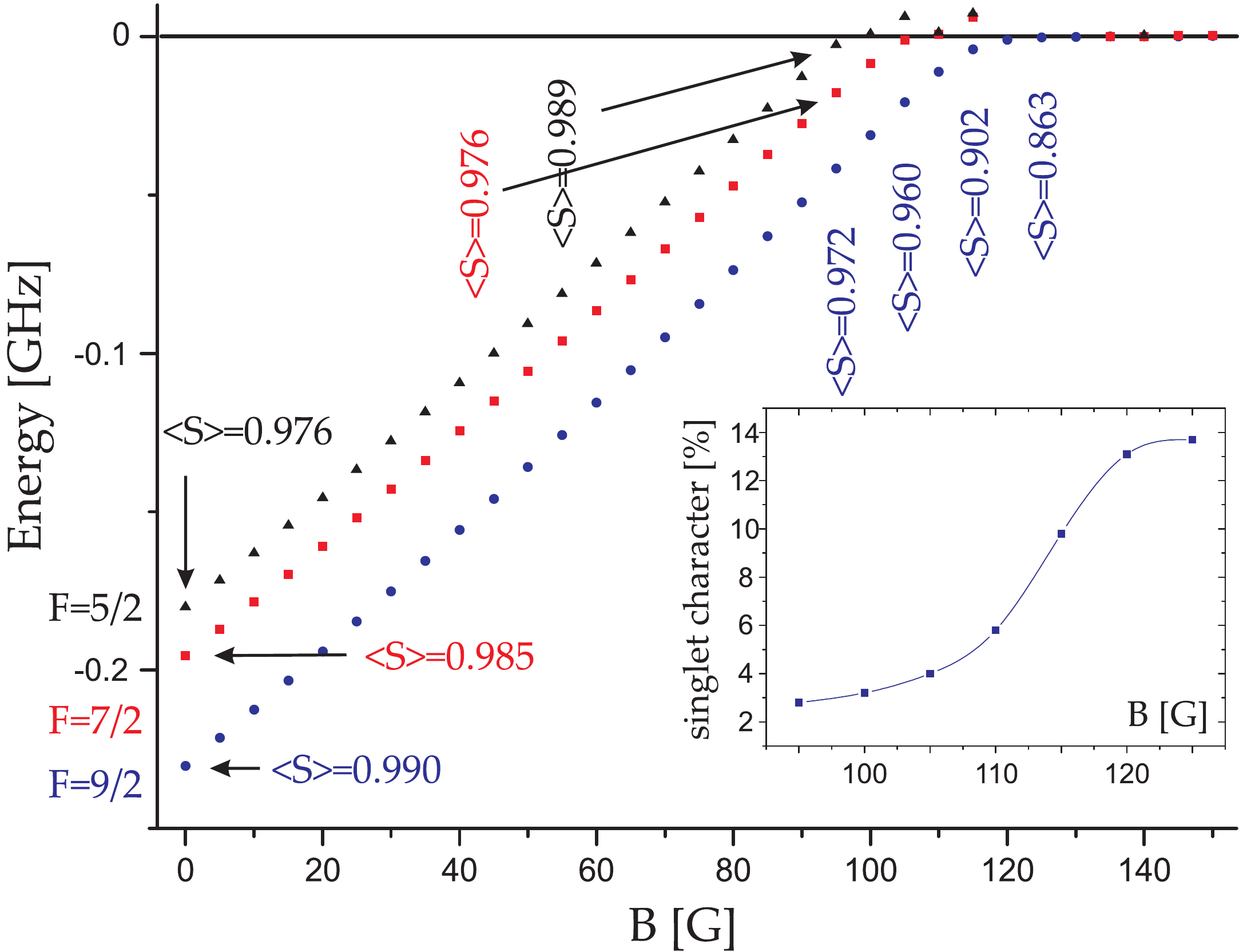}
\caption{(color online) Binding energies of $^{23}$Na$^{40}$K for the total quantum number $\mathcal{M} = - \nicefrac{3}{2}$ as function of magnetic field leading to $s$-wave resonances for the entrance channel (1,1)$_{Na}$ + (9/2,-5/2)$_{K}$. For various magnetic fields the expectation value $\langle S \rangle$ of the electronic spin is given. Inset: Singlet character as a function of $B$ for the resonance situated at 136 Gauss. For the sharp resonances at lower fields, only a subtle change in spin character is observed.}
\label{fig2}
\end{figure}

\subsection{Feshbach resonances and Feshbach spectrum}

For the nature of the Feshbach molecular wavefunction, we perform coupled-channel calculations of the ground state levels, taking into account the coupling of \Xstate~and \astate~due to hyperfine interaction and the Zeeman effect at elevated magnetic fields applied for the creation of Feshbach molecules. The calculations in \cite{Tiemann} reproduce the $s$-wave resonances measured in~\cite{Zwierlein1} within an uncertainty of 0.1 G regarding their position and extend the analysis of the aforementioned article by an accurate description of the $p$-wave multiplet structures. The atom pair states are fully described by the atomic basis set $\Ket{(i,s,f,m)_{\mathrm{Na}};(i,s,f,m)_{\mathrm{K}},F,m_F}$ simplified to $\Ket{f_{\mathrm{Na}},m_{\mathrm{Na}},f_{\mathrm{K}},m_{\mathrm{K}},F,m_F}$, where $s_{A} (i_{A})$ is the electron (nuclear) spin, $f_{A}$ the total angular momentum and $m_{A}$ its projection onto the space fixed axis of atom A, and $F, m_F$ the total angular momentum and its projection of the system excluding rotation. This basis is called Hund's case (e). The number of channels is given by the number of possible projections of the individual angular momenta onto the space fixed axis equating to the same total magnetic quantum number. The following discussion concentrates on the fermionic molecule $^{23}$Na$^{40}$K, and assumes that the initial atomic states have been converted to molecules possessing a total magnetic quantum number of $\mathcal{M} = - \nicefrac{3}{2}$, for which the total number of channels is sixteen for a rotational state $l=0$, i.e. $\mathcal{M} =m_F$. 
	
\begin{figure}
\centering
\includegraphics*[width=0.95\columnwidth]{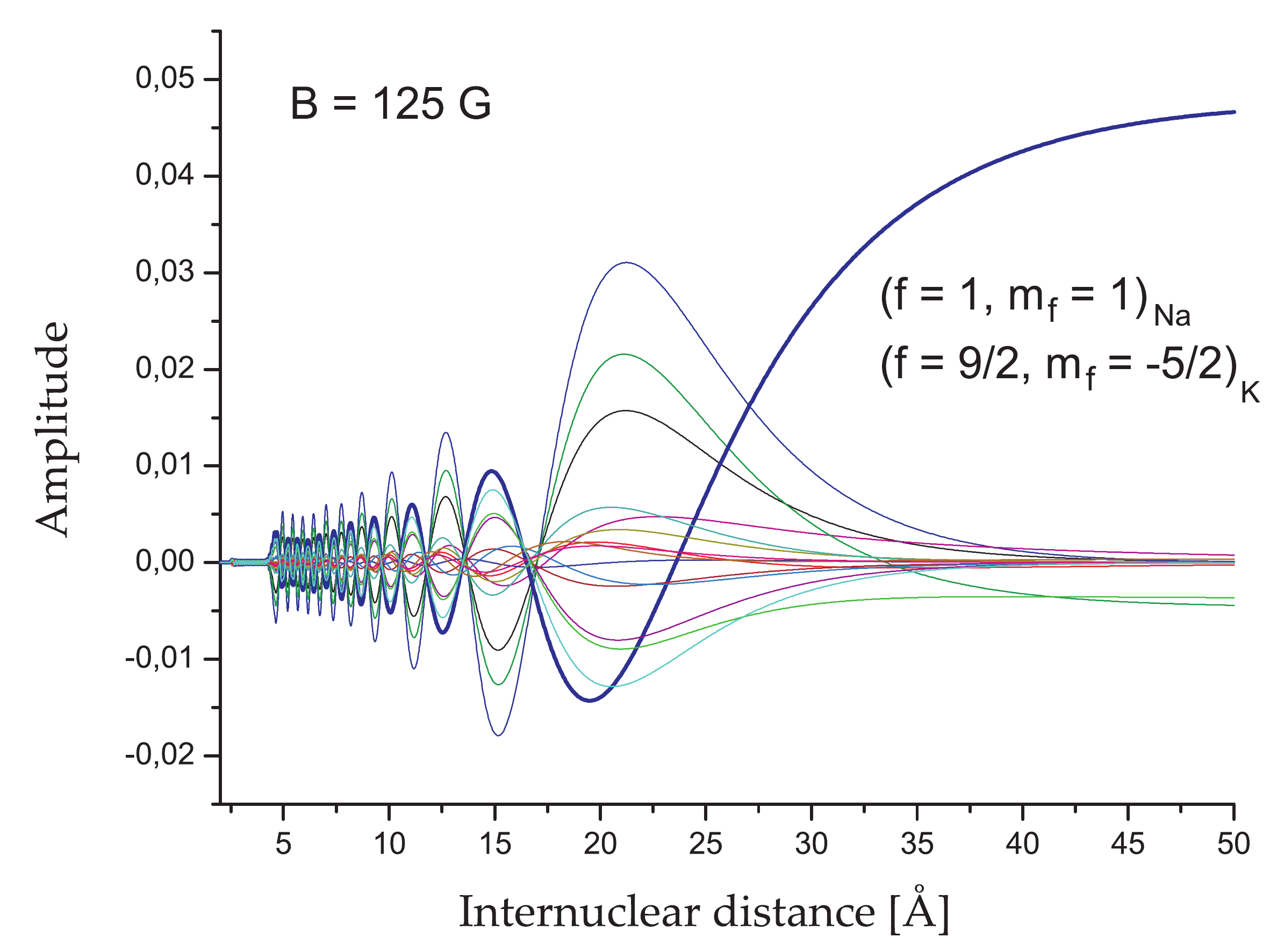}
\caption{(color online) Multi-component wavefunction for the bound level at 125 Gauss, projected onto the Hund's case (e) basis. The wavefunction displays a distinct open channel fraction by the large amplitude at large internuclear distance. This component is named by the appropriate quantum numbers.}
\label{fig125e}
\end{figure}

The calculated binding energies for  $\mathcal{M} = - \nicefrac{3}{2}$ are shown in Fig. \ref{fig2} below the asymptote of the atom pair (1,1)$_{Na}$ + (9/2,-5/2)$_{K}$ and give rise to the three $s$-wave resonances situated and observed in~\cite{Zwierlein1}  between 96 and 138 Gauss.

\subsection{ Spin character of the Feshbach state}

Having obtained the scattering multi-channel wavefunction, we are now ready to inspect its spin composition as a function of the applied magnetic field. The closed channel state will be dominantly a bound state of the \astate potential, hence notable singlet wavefunction amplitude can only arise out of the open channel. As the resonance width can be linked to the coupling strength between the collision channels, the wide $B=138$ G Feshbach resonance seems a promising candidate for generating singlet character in the Feshbach state. In~\cite{Zwierlein2}, it was shown that the singlet admixing increases as the magnetic field approaches the resonance. This is confirmed by our calculations, where we evaluate the expectation value of the total spin operator as specified in Fig. \ref{fig2}. A comparison of the three $s$-wave resonances shows that the spin character only changes significantly for the broad resonance, saturating at the open channel singlet admixture near the resonance. The latter can be calculated by performing angular momentum recoupling into a representation in which the total electronic and nuclear spin quantum numbers $S,I$ are defined. The unitary transformation from such basis $\Ket{S,I,F,\mathcal{M}}$  to the desired atomic basis of the entrance channels $\Ket{f_{\mathrm{Na}},m_{\mathrm{Na}},f_{\mathrm{K}},m_{\mathrm{K}},\mathcal{M}}$ then reads
\begin{align}
&\Ket{f_{\mathrm{Na}},m_{\mathrm{Na}},f_{\mathrm{K}},m_{\mathrm{K}},\mathcal{M}} = \notag \\ &\sum_{F,S,I} \braket{f_{\mathrm{Na}}f_{\mathrm{K}}|S,I} \braket{m_{\mathrm{Na}},m_{\mathrm{K}}|F,\mathcal{M}} \Ket{S,I,F,\mathcal{M}}  
\end{align}
with the transformation coefficient applying the Wigner 9j symbol:
\begin{align}\braket{f_{\mathrm{Na}}f_{\mathrm{K}}|S,I} =  \begin{Bmatrix} s_{\mathrm{Na}} & i_{\mathrm{Na}} & f_{\mathrm{Na}}\\ s_{\mathrm{K}} & i_{\mathrm{K}} & f_{\mathrm{K}} \\ S & I & F \end{Bmatrix} \notag \\
 \sqrt{(2f_{\mathrm{Na}}+1)(2f_{\mathrm{K}}+1)(2I+1)(2S+1)}
\end{align}
and the Clebsch-Gordan coefficient $\braket{m_{\mathrm{Na}},m_{\mathrm{K}}|F,\mathcal{M}}$  projects onto the magnetic submanifold. Summing over all possible cases for $\mathcal{M}=-3/2$ gives a singlet $(S = 0)$ fraction of $\sim 18\%$ for the open channel $\mathcal{M} = - \nicefrac{3}{2}$~at the lowest asymptote, roughly matching the saturation behavior by 14\% shown in the inset of Fig. \ref{fig2}; the difference is induced by the competition between hyperfine and Zeeman coupling at 125 G, when $f_{Na}$ and $f_{K}$ will no longer be exact quantum numbers. As one shifts away from the resonance to lower fields, the molecular state converges towards an almost pure triplet state. 

\subsection{Feshbach molecular wavefunction and its magnetic field dependence}

We now inspect the wavefunction, as it comprises more information than the spin character. Near the Feshbach resonance, one has to distinguish between 'atomic' open-closed channel mixing on one side, and 'molecular' singlet-triplet mixing on the other side. To provide an intuitive insight into both perspectives, we utilize two different representations of the total multi-channel wavefunction. At large internuclear distances, Hund's case (e) is a convenient choice of a basis set, as the coupling to the molecular axis plays a secondary role only. By projecting the wavefunction onto the atomic basis $\Ket{f_{\mathrm{Na}},m_{\mathrm{Na}},f_{\mathrm{K}},m_{\mathrm{K}},F}$, one obtains direct access to the open and closed channel character of the scattering wavefunction, as seen in Fig. \ref{fig125e}. At a magnetic field of 125 Gauss about 13 G below the very broad resonance, the bound level is at -250 kHz and strong coupling to the open channel persists, which is highlighted by the large amplitude of one single basis vector, labeled by quantum numbers. As the open channel contains a considerable fraction of singlet character, this directly translates into an increased singlet-triplet ratio for the Feshbach state. At first glance, one could interpret a high singlet admixing as being beneficial for the desired process for ground state transfer, because one could directly couple to a more or less pure singlet intermediate state. Yet components going over to open channels possess significant wavefunction amplitude only at large distances, which will not contribute to the transition dipole matrix element in eq. \eqref{eq:dip} due to vanishing vibrational overlap. Hence the wavefunction amplitudes and their state character localized at smaller internuclear distances are of great interest. 

\begin{figure}
\centering
\includegraphics*[width=0.95\columnwidth]{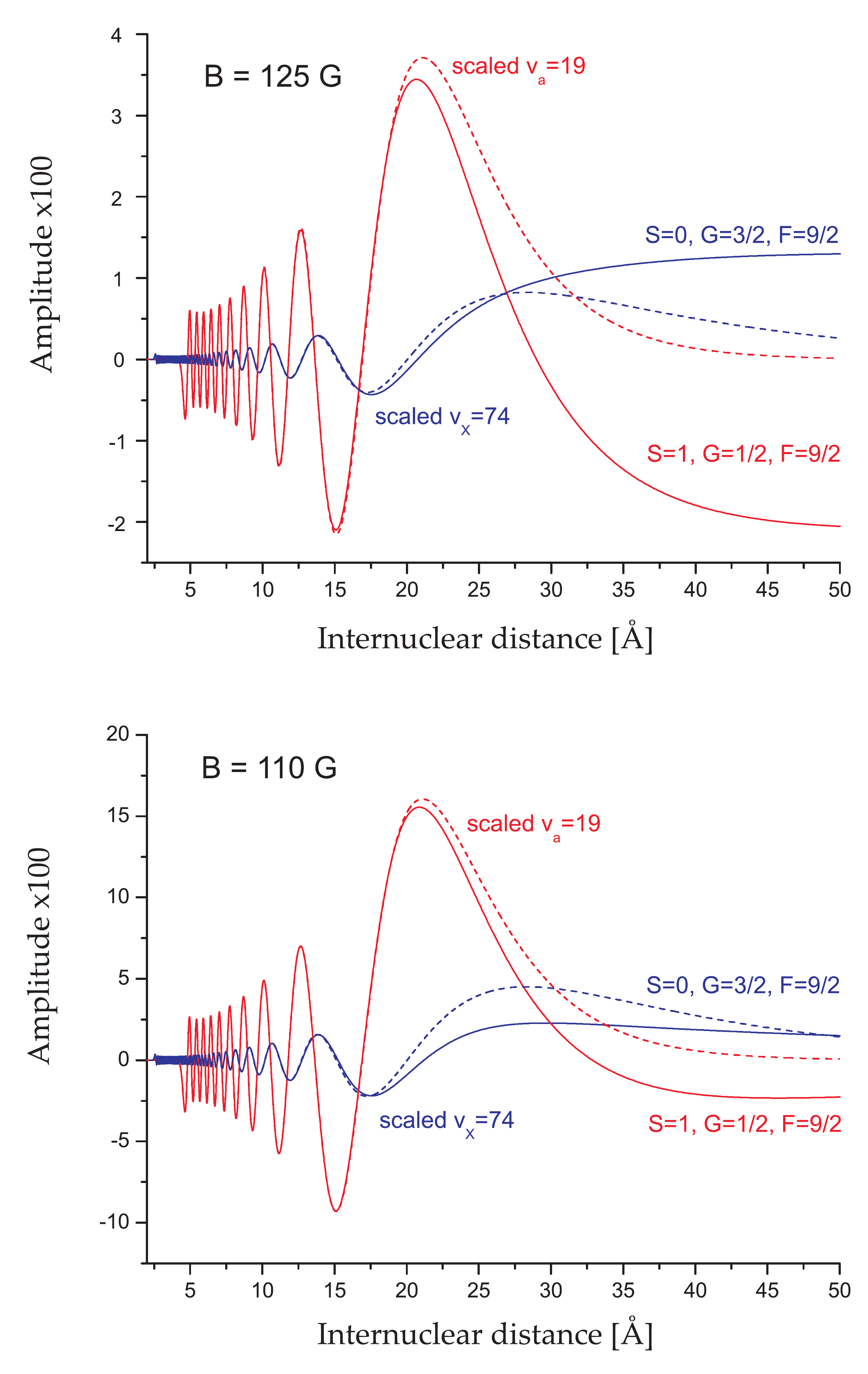}
\caption{(color online) Multi-component wavefunction of the bound level at 125 (upper graph) and 110 (lower graph) Gauss, projected onto the Hund's case (b) basis. For clarity, all but the strongest triplet and singlet channels were removed. Note the different scaling on the amplitude axis. As dashed lines the vibrational wavefunctions of the pure singlet and triplet states are given for the least bound level, scaled to the amplitude of the appropriate inner parts of the multichannel wavefunction.}
\label{figb}
\end{figure}

The Hund's case (e) representation is not suitable for directly identifying the singlet and triplet admixture, as the total spin is not appearing as a quantum number in the basis vector. For that reason we transform the Feshbach wavefunction into a state basis in which the total spin is used, namely Hund's case (b). Here, the angular momentum coupling, neglecting in our special case the molecular rotation, gives rise to a set of quantum numbers $\Ket{S,G,F,M}$, where G is obtained by coupling the total spin S with the nuclear spin of sodium (note that the sodium hyperfine splitting is larger by a factor of 1.4 than for \K).
The resulting projections of the wavefunction onto the basis (b) are shown in Fig. \ref{figb} for a magnetic field of 125 Gauss (upper graph, binding energy 250 kHz$\times $h) and 110 Gauss (lower graph binding energy 11 MHz$\times $h), respectively. For reasons of clarity and comprehensibility, we removed all but the strongest triplet and singlet channel contributions.  From Fig. \ref{figb} it is evident that one benefits from  changing the magnetic field from 125 to 110 G, as it leads to compacting the Feshbach wavefunction at internuclear distances $r < 20$ \AA. Both triplet and singlet amplitudes are enhanced in this range at 110 Gauss. To quantify this, we calculated the single-channel wavefunctions for the last bound levels of the \Xstate~ and \astate~potentials, v$_X$=74 and v$_a$=19, respectively. At internuclear distances drawing near the chemical region, the amplitude from the continuum coupling will be damped out and the individual channel wavefunctions will change over to the corresponding bound state wavefunctions. We match their amplitude for $r \lesssim 10 $\AA~(see Fig. \ref{figb}), resulting in scaling factors $\beta_{\mathrm{a}}(B),\beta_{\mathrm{X}}(B)$ for the unperturbed single-channel wavefunctions. These scaling factors can be interpreted as the amplitude gain or loss for the Franck-Condon overlap. For the two magnetic fields shown in \ref{figb}, this gives a ratio $\beta_{\mathrm{X}}(110)/\beta_{\mathrm{X}}(125) = 5.48$ for the strongest singlet and $\beta_{\mathrm{a}}(110)/\beta_{\mathrm{a}}(125) = 4.33$ for the strongest triplet channel. Evaluating Feshbach transition dipole matrix elements, this gain in $\beta_{a,X}$ directly translates into a transition probability gain of one full magnitude.

Concluding, the rise in singlet character close to the resonance field is accompanied by significant total amplitude loss in the inner part. This behavior cannot be revealed by inspecting the spin character, as it cloaks such details due to the integration procedure. Instead, one has to look directly in the multi-channel wavefunction, which naturally contains detailed information. Note that this behavior is depending on the resonance that one works with (e.g. the sharp resonance at 96 G displays slightly increasing singlet character as one moves from the resonance). For each resonance, there will be a magnetic field which optimizes the transfer to the molecular ground state via a selected intermediate state. This reflects the competition of singlet amplitude admixed by the open channel coupling and the increase in amplitude of the desired component in the inner region of the wavefunction. 

\section{Intermediate states}\label{intermediate}

The intermediate state will be one of the eigenstates of the Hamiltonian
\begin{equation}
\mathcal{H} = H_{0} + H_{\mathrm{Int.}}.
\end{equation}
Here, $H_{0}$ contains the kinetic energy and diabatic potential operators of a specific molecular state manifold. The generally used manifold corresponds to either one of the two lowest electronic excitations Na(3$p$)+K(4$s$) or Na(3$s$)+K(4$p$), and the spin-orbit coupling will be the dominant interaction in $H_\mathrm{Int.}$. Since the atomic spin-orbit coupling constant  for the lowest p-state of K is about a factor of 3 larger than the corresponding coupling for the lowest p-state of Na ($\xi_{K,[Na]} = 19.24 [5.73]\,\mathrm{cm}^{-1}$), we choose the states asymptotically converging to  K(4$p$) for our analysis, which is also the energetically lowest one of both sets mentioned above. Note that we want to populate $\Ket{\psi_{X}}$ from the intermediate states. $\Ket{\psi_{X}}$ is symmetric under parity inversion for the rotational state J=0 demanding for levels with odd parity for the excited states. For the asymptote under consideration, this then gives five molecular state vectors. For state labeling, we use the $^{2S+1}\Lambda^{\pm}_{\Omega}$ symmetry (Hund's case (a)), where $\Lambda$ ($\Omega$) gives the projection of the electron orbital (total) angular momentum along the internuclear axis. The manifold under consideration then comprises two states $^{1}\Sigma^{+}_{0^{+}}$,$^{3}\Pi_{0^{+}}$ sharing $\Omega = 0^{+}$ and three states $^{3}\Sigma_{1}^+$,$^{1}\Pi_{1}$,$^{3}\Pi_{1}$ sharing $\Omega = 1$. 

The interaction Hamiltonian $H_\mathrm{Int}$ can be divided into hyperfine and  spin-orbit interaction. Despite playing a pivotal role for the ground states, the hyperfine interaction is neglected in the following discussion, as it will be overshadowed by the other interactions and does not change the overall picture significantly. We also set $B = 0$ and therefore do not discuss magnetic field effects.  The spin-orbit interaction couples states possessing the same value of $\Omega$ ($\Delta\Omega=0$).  But there is still the Coriolis interaction from $H_0$ coupling states with $\Delta\Omega=\pm1$. $\Omega=2$ plays no role because such states do not posses J'=1 levels.

\subsection{Modeling excited state molecular potentials}
The reliability of our calculations will be dictated by the accuracy with which the applied PECs and their interaction reflect reality. \textit{Ab initio} curves provided by quantum chemistry calculations ~\cite{abinitio} give a complete overall description of the PEC, which, however, is accompanied by lower total accuracy. The desired states have also been investigated by means of molecular spectroscopy~\cite{Stolyarov,Kato,Barrow}. Out of these measurements, spectroscopic constants can be extracted and RKR curves constructed. The resulting molecular potentials reproduce the bound molecular states of the chemical region with a quality which considerably exceeds the \textit{ab initio} approach. Yet the RKR treatment is only applicable for the region in which the molecular states have been explored experimentally. If RKR potentials are extrapolated beyond the spectroscopically investigated range, the potential slope errors become large. It is therefore advised  to use these RKR potentials only to describe the potential minimum part. The short range part of the PECs can be modeled by a repulsive wall involving a high inverse power in internuclear distance $R$. At internuclear distances larger than the LeRoy radius~\cite{LeRoy}
\begin{equation}
R \geq 2 \left\{ \sqrt{\langle r_{\mathrm{Na}}^{2} \rangle} + \sqrt{\langle r_{\mathrm{K}}^{2} \rangle} \right\},
\end{equation}
with $\langle r_{i}^{2}\rangle$ being the expectation value of the squared radius of the outermost electron on the i-th atom, the atoms of the dimer can be considered detached and the long-range description takes over. This is given as an inverse power series involving the individual dispersion coefficients, where high quality theoretical values are tabulated in the literature~\cite{Marinescu,Derevianko}. The LeRoy radius for NaK is reported as 10.8 \AA~\cite{Kato}, yet the only RKR curve fully covering the region until that value is the B$^{1}\Pi$ ~\cite{Kato} one, thus it implies that one has to borrow an  \textit{ab initio} shape in order to bridge the part between RKR and long-range description. To avoid mixing theoretical and semi-empirical descriptions, we decided to follow a different path described below, taking the \textit{ab initio} curves of ~\cite{aymar} and refine them by using our fully explored ground state potentials. The RKR potentials~\cite{Stolyarov,Kato,Barrow} are then employed together with spectroscopic data as a crosscheck for the spectroscopically known regions. 

\begin{figure}
\centering
\includegraphics*[width=0.95\columnwidth]{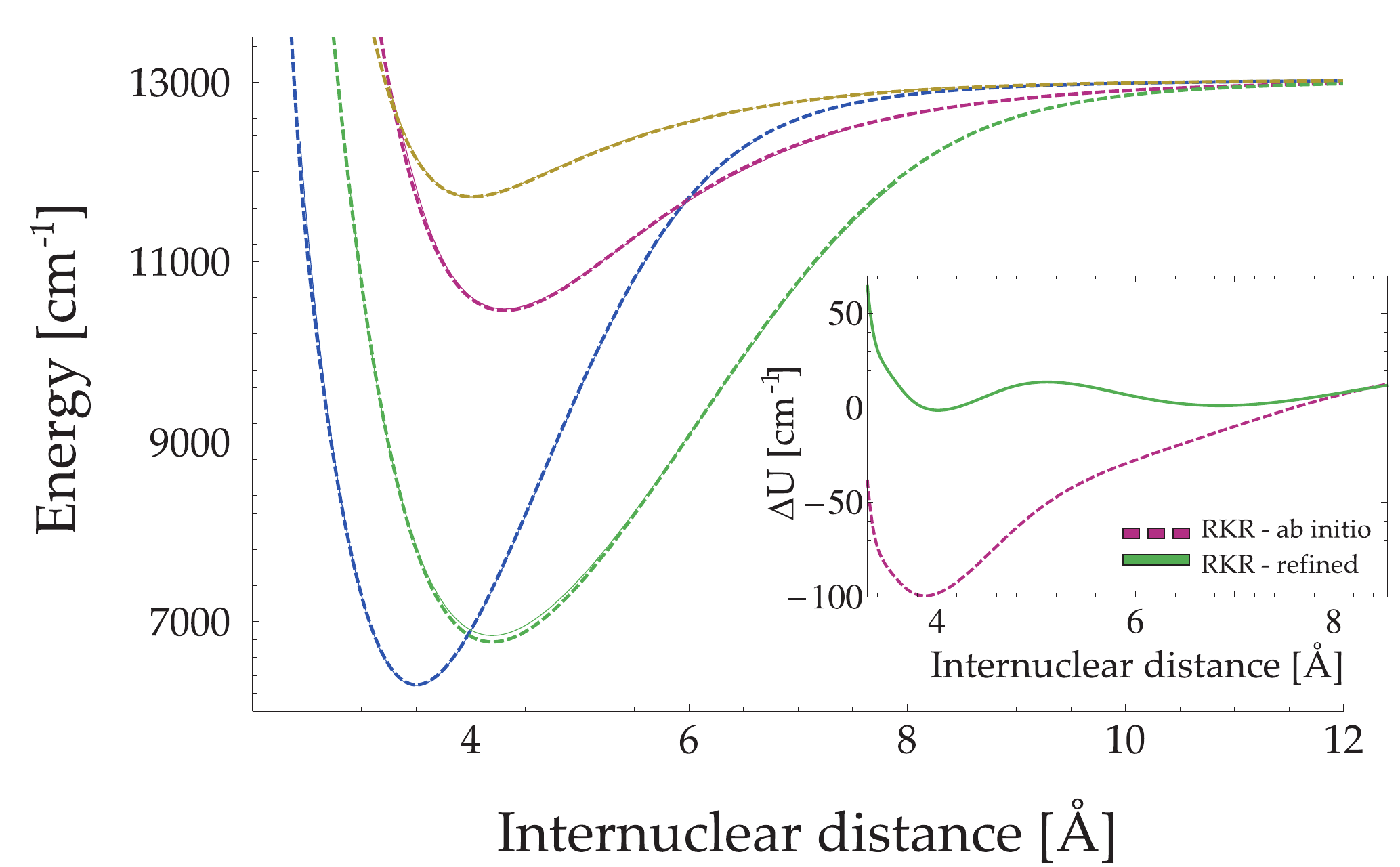}
\caption{(Color online) Diabatic PECs of the Na$(3s)$ + K$(4p)$ manifold used in the study. The energy zero is chosen at the atomic ground state asymptote. Shown are the diabatic PECs applied in this section (dashed) as well as RKR curves (solid) for comparison. Inset: Potential difference between the B$^{1} \Pi$ RKR potential, the raw \textit{ab initio} curve (dashed line) and the refined curve (solid line), respectively.}
\label{figpashov}
\end{figure}

\textit{Ab initio} calculations rely on approximations which in the end will over- or underestimate certain facets of the potentials as for example the well depth of the potentials. It is noted that these errors are systematic, e.g. $R$-dependent errors in the theoretical curves are mostly generated out of basis-set superposition errors, so they will in general affect all calculated curves in similar fashion. Possessing reliable experimental data on the ground states, we can compare those to the \textit{ab initio} ground state PECs. Out of this we obtain information regarding systematic deviations in the \textit{ab initio} calculations, which we then apply to correct the excited state potentials~\cite{NaRb2,NaRb1}. We further refine the potentials by performing our calculations for rotational progressions which we have studied experimentally. With this help, we extract common energetic offsets as well as $\Delta B_{i} = B_{i,\text{calc.}} - B_{i,\text{Exp.}}$, where $B_i$ is the rotational constant of the i-th molecular state. Assuming $B \sim \nicefrac{1}{(R_{eq.}^{2})}$, we balance this difference $\Delta$ by slightly shifting the equilibrium distance. This whole procedure considerably improves the quality of the curves, as shown exemplary in the inset of Fig. \ref{figpashov} for the B$^{1} \Pi$ potential. It is reliable for the singlet states, as the region, where the ground state refinements are applied, is well covered by our spectroscopic studies. For the triplet states, one has to employ refinements through the corresponding a$^{3}\Sigma^{+}$ state, which gives  good results for $R > 4.55$  \AA~up to and beyond the LeRoy radius. For smaller internuclear distances, the short-range description could be used from the ground state, but already small differences in the short-range parameters could lead to massive discrepancies due to the high inverse powers involved, and the refinement procedure would actually worsen the initial potential. To circumvent artifacts arising of such modifications, we smoothly connect the refined part at $R = 4.65 \text{\AA}$ with the respective RKR parts. The final PECs used in our calculation are shown in Fig. \ref{figpashov} (dashed lines) together with the RKR curves obtained from~\cite{Stolyarov,Kato,Barrow} (solid lines), showing satisfactory agreement. 

\begin{figure}
\centering
\includegraphics*[width=0.95\columnwidth]{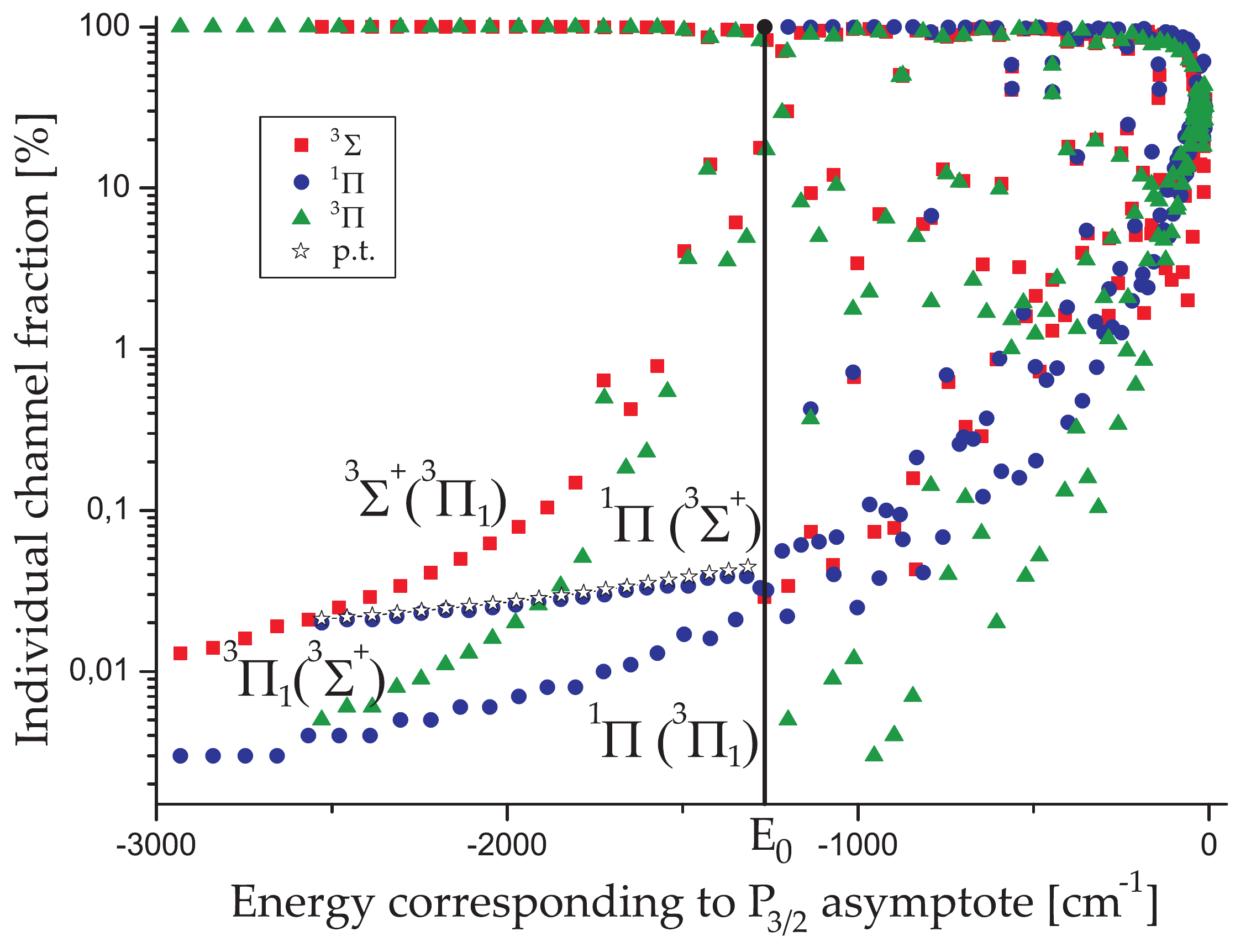}
\caption{(Color online) Integrated square of the individual $\Omega = 1$ channel wavefunctions at each eigenenergy, together with a $J=1$ perturbative model for the $^{1}\Pi (^{3} \Sigma^{+})$ case (open stars). A label of the form $^{3}\Sigma^+(^{3}\Pi)$ marks the $^{3}\Sigma^+$ fraction within a dominant $^{3}\Pi$ state. The vertical line at $E_0$ is positioned at the first state with dominant $^{1}\Pi$ character.}
\label{figpert}
\end{figure}

\subsection{Spin character of excited molecular states}
Similar to our ground state discussion, we begin by inspecting the spin character of the individual eigenstates. For finding the eigenstates the molecular state vectors in (a) are unitarily transformed into the Hund's case (e) basis, as the spin-orbit operator, mainly H$_{int}$, is diagonal in this representation and H$_0$ is also transformed to case (e). A diagonalization procedure then gives rise to a set of eigenstates and -energies for each total angular momentum $J$, spanning 7000\,cm$^{-1}$ for the full depth of b$^{3}\Pi$.  Projecting the eigenstates onto the Hund's case (a) state vectors and integrating their squared amplitudes yields the fractions of the individual channels. This is shown in Fig. \ref{figpert} for $\Omega = 1$ and a frequency window of 3000\,cm$^{-1}$ below the dissociation limit $3^2S_{1/2}$(Na)$+4^2P_{3/2}$(K). The symbol $^{3}\Sigma^{+}(^{3}\Pi)$ marks the channel fraction of $^{3}\Sigma^{+}$ within a dominant $^{3}\Pi$ state and correspondingly all other symbols. Being interested in the singlet-triplet admixture, we focus on the $^{1}\Pi-{}^3\Pi$ and $^{1}\Pi-{}^3\Sigma^{+}$ coupling because this promises to have a strong Stokes transition from $B^{1}\Pi$ to $X^{1}\Sigma^{+}$. The resulting structure can be briefly divided into two regions, which are naturally split at the energy $E_{0}$ of the lowest level of $B^{1}\Pi$, being indicated in Fig. \ref{figpert} by a black vertical line. 

The region lying energetically below $E_{0}$ can be classified as a perturbative region. The eigenstates display $>99\%$ character of either the $^{3}\Sigma^{+}$ (red squares) or $^{3}\Pi_{1}$ (green triangles) channel in the upper line of Fig. 6. Note that this corresponds to a distinct molecular structure, where the assignment of measured energies to vibrationally bound levels is straightforward.  In this area, $^{1}\Pi$ (blue dots) fraction is generated in almost pure $^{3}\Sigma^{+}$ states via spin-orbit coupling and in almost pure $^{3}\Pi_{1}$ states via higher order effects, as the direct coupling matrix element between $^{1}\Pi$ and $^{3}\Pi_{1}$ is strictly zero. The spin-orbit induced singlet admixture in $^{3}\Sigma^{+}$ states becomes $0.02-0.04 \%$

This result can be understood by a simple perturbative approach.  Let us denote the k-th eigenstate of $^{1}\Pi$ as $\Ket{\psi^{^{1}\Pi}_{k}}$ and its eigenenergy as $E_{k}$ and similarly the n-th bound $^{3}\Sigma^{+}$ level by $\Ket{\psi^{^{3}\Sigma}_{n}}$ and $E_{n}$. Approximating the spin-orbit operator by its atomic coupling constant $\xi_K$, the first order state correction of the almost pure triplet manifold becomes
\begin{equation}
\sum_{k} \frac{\braket{\psi^{^{1}\Pi}_{k}|\hat{\xi}| \psi^{^{3}\Sigma}_{n}}}{E_{k} - E_{n}} \Ket{\psi^{^{1}\Pi}_{k}} \approx \xi_{K} \times \sum_{k} \dfrac{\mathrm{FCF}(k,n)}{\Delta E_{k,n}}\Ket{\psi^{^{1}\Pi}_{k}}
\label{eq:pert}
\end{equation}
where the problem reduces to a calculation of the vibrational overlap FCF between the states of interest and their respective energy difference $\Delta E_{k,n}$. Summing the squared coefficients of eq. \eqref{eq:pert} results in the singlet fraction of the $n$-th triplet state, which is also displayed in Fig. \ref{figpert} (open stars), and the good agreement with the simulation emphasizes the perturbative character of the admixing. Note that the absolute amount is slightly overestimated by the perturbative approach. We mention that the dominant contribution in  eq. \eqref{eq:pert} to the interaction comes from the k-th vibrational $^1\Pi$ level sharing considerable ($42 \%$ mean) wave function overlap with the $n=k+2$-th vibrational $^3\Sigma^{+}$ level, which originates accidentally from the relative forms of the potentials and their relative positions in R.  

For completeness, let us also discuss the $^{3}\Sigma^{+}-^{3}\Pi$ interaction. Far below $E_0$, the eigenstates have either dominantly  $^{3}\Sigma^{+}$ or $^{3}\Pi$ character. However, the interaction features a frequency window of around 400\,cm$^{-1}$ below $E_{0}$, in which the assignment to an experimental observation becomes difficult because the admixtures of either $^{3}\Sigma^{+}$ ($^{3}\Pi$) character to $^{3}\Pi$ ($^{3}\Sigma^{+}$) exceed $1\%$. As the eigenenergies approach around $E \approx-1200\,\mathrm{cm}^{-1}$, the mixed amount rapidly increases to $30 \%$ and decreases subsequently, displaying a resonant behavior. In this energy interval the two potentials $^{3}\Sigma^{+}$ and $^{3}\Pi$ cross each other. The vibrationally averaged interaction parameter then largely exceeds the vanishing frequency difference, and the crossing resonance occurs. The appearance of this structure near $E_{0}$ is accidental. 

At and above $E_{0}$, the structure in Fig. \ref{figpert} becomes complicated, and work in this area will be accompanied by increased spectroscopic effort. We classify this as the resonant region, because the presence of vibrationally bound $^{1}\Pi$ states is accompanied by levels of other electronic states. These resonances are qualitatively different from the crossing resonances discussed before. They appear due to energetically nearly degenerate singlet and triplet rovibrational levels $(\Delta E_{n,k} \lesssim 5\,\mathrm{cm}^{-1})$, when the interaction parameter will be larger than the energy difference and any perturbative treatment will break down. The resonantly interacting states share no similarities concerning relative vibrational quantum numbers, and no simple pattern can be identified in the spectrum. 

We note that such accidentally resonant structures also appear below the first $^{1} \Pi$ state and belong to $^{1}\Sigma^{+}-{}^3\Pi_{0}$ states coupling with $\Omega = 0^{+}$, which have been removed from Fig. \ref{figpert} for clarity. Also a  perturbative region admixing $^1\Sigma^{+}$ character to dominantly $^3\Pi_{0}$ states exists just below the potential minimum of the $^{1}\Sigma^{+}$ state about 6000\,cm$^{-1}$ below the asymptote, and is almost overlayed by the $^{1}\Sigma^{+}-{}^3\Pi_{0}$ crossing resonance. As the wavefunctions involved in this low energy region only spread over a small internuclear distance range, the overlap with the Feshbach molecular wavefunction  will be small and two-photon Raman transfer (STIRAP) not efficient. The general two-photon process will be quantified in the section V of this article.

\subsection{Discussion}
 We shortly discuss the suitability of the found structures (second order admixing, crossing resonances and degenerate resonances) for the two-photon process and comment on their robustness to changes in the PECs. Shifting the potential curves in terms of total energy or equilibrium distance will shift the crossing resonances accordingly to the new crossing position, but will not alter their fundamental structure. On the other hand, the direct resonances occurring in near-degenerate states will react highly sensitive to any potential change due to their accidental nature. Despite the good agreement of our PECs with experimental data, it puts large uncertainties in the actual mixing value of such a resonance. In fact, by changing the PEC parameters slightly, some resonances will vanish completely and others appear. Our model is therefore not reliable for giving quantitatively exact predictions of the positions and values of the resonant mixtures. To fill this gap in our model, we are presently complementing our description by spectroscopic work applying molecular beams for sufficient resolution. A first example of such cases is given below in section \ref{sec:STIRAP}. 

The second order admixed states in the perturbative region will present the most robust situation. They will be largely unaffected by energetic offsets, as it only varies the energy denominator in eq. \ref{eq:pert}, and a mismatch by 1\,cm$^{-1}$ will change $\Delta E_{n,k}$ by less than 0.1 percent. Also slight radial mismatches in the turning points will only introduce minor corrections due to the integral nature of the $^{3}\Sigma^{+}$ state perturbation by the $^{1}\Pi$ manifold. Together with the relative spectroscopic ease which awaits one at such a perturbative level, this highlights the perturbative region as being a good candidate for the desired STIRAP transfer. Yet the absolute value of the singlet spin character might cast doubt on the suitability of these states for the two-photon process. For the KRb analogue, the used level was reported to possess a singlet character of $0.2\%$ \cite{kotochigova}. By just comparing the SO coupling strengths, one would expect the NaK case to have a factor $(\xi_{Rb}/\xi_{K})^{2} \approx 17$ lower admixture. This is partially compensated for due to the earlier discussed potential shapes, which favor the NaK case. The NaK admixture of around $0.04 \%$, a typical value from Fig. \ref{figpert}, is still a factor of 5 lower than the KRb one. For a definite statement, one has again to inspect closely the wavefunctions and calculate transition matrix elements, which is quantified below.

\section{Two-photon process}
\label{sec:STIRAP}

In the previous sections, we provided a full understanding of ground and excited state molecular potentials and the resulting molecular levels involved in the envisaged two-photon process from Feshbach molecules to rovibrational ground state molecules in the $X^1\Sigma^{+}$ potential. We are now ready to quantify the two-photon transition matrix element $d_{(\mathrm{F}\rightarrow\mathrm{X})}$ given by equation \eqref{eq:dip}. For the Feshbach state we use the broad $F = \nicefrac{9}{2}$ resonance appearing for $\mathcal{M} = -\nicefrac{3}{2}$. Decomposing the Feshbach molecular state vector yields the individual channel contributions. In our calculations, we take only the two strongest channels into account.  The Feshbach molecular state can then be approximated by

\begin{align}
\Ket{\psi_{F}(B)} \approx \; \beta_{\mathrm{X}}(B) \Ket{X^{1} \Sigma^{+} (v = 74, J = 0)}& \\
 + \beta_{\mathrm{a}}(B) \Ket{a^{3} \Sigma^{+} (v = 19, N = 0)}& \\
 + \text{amplitude of other hyperfine channels} &
\end{align}
where $\beta_i$ ($i=a,X$) are the scaling factors obtained in the matching procedure of section \ref{groundstate}. Note that by cutting off the other channels, interference effects of those contributions are ignored, and only interference of the strongest channels with each other is considered. The neglected amplitude would gain importance when evaluating the hyperfine structure for the excited states.

We calculate the transition dipole matrix elements from the Feshbach molecular state and the X$^1\Sigma^+(v=0,J=0)$ state to  all $j = 1,\ldots, 683$ electronically excited bound eigenstates obtained up to 10 cm$^{-1}$ below the atomic asymptote $P_{\nicefrac{3}{2}}$ of Potassium. They formally read $| \psi_{\text{int.,j}}\rangle = \sum_{i} c_{i,j} \Ket{i}$, with $i = \left\{{}^1\Sigma^+,{}^3\Sigma_1^+,{}^1\Pi,{}^3\Pi_{0},{}^3\Pi_{1}\right\} $, where the sums of the squared coefficients give the channel fractions shown in Figure \ref{figpert} of section \ref{intermediate}. The singlet and triplet electronic transition dipole moments are taken from~\cite{aymar}.

\begin{figure}
\centering
\includegraphics*[width=0.95\columnwidth]{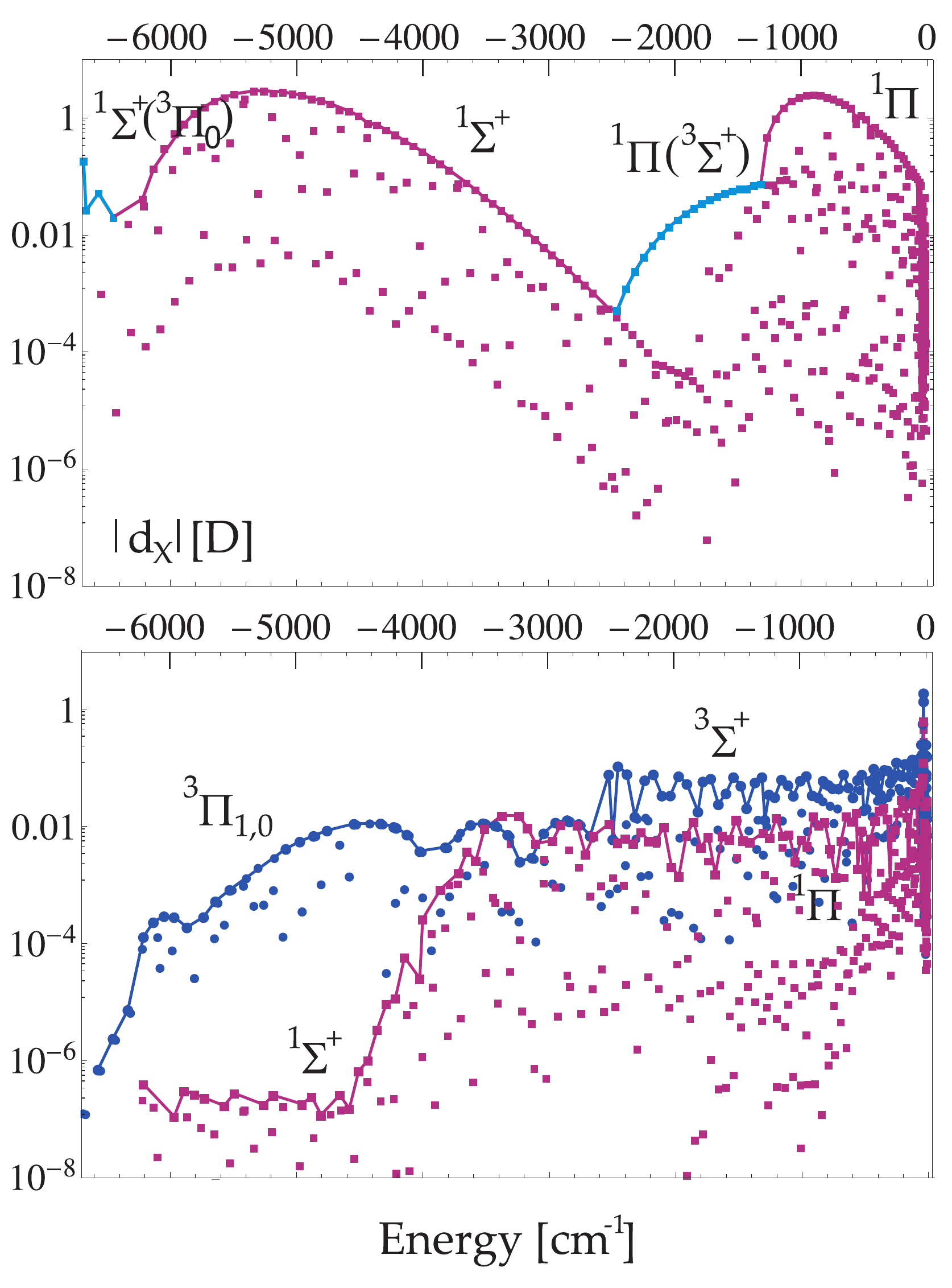}
\caption{(Color online) Absolute values of the Stokes ($\Ket{i}\rightarrow\Ket{X}$ upper graph) and Pump ($\Ket{F}\rightarrow\Ket{i}$ lower graph, decomposed into the singlet (red squares) and triplet (blue circles) part) laser transition dipole matrix elements for all intermediate states. The energy is given with respect to the $\mathrm{P}_{\nicefrac{3}{2}}$ dissociation limit.} 
\label{x-a-leg}
\end{figure}

 The resulting absolute values $|\mathrm{d}_{X}|$ of the dipole matrix elements from the ground state \Xstate$(v=0,J=0)$ to the intermediate states and $|\mathrm{d}_{a}|$ from the Feshbach state to the intermediate states are shown in Figure \ref{x-a-leg} as a function of the intermediate state eigenenergy for a magnetic field of  $B = 110$ G for all states with $\Omega = 0^{+}$ and $\Omega = 1$. In the upper graph, locally strongest transitions to singlet (triplet) dominant eigenstates have been connected by a red (blue) line, serving as a guide to the eye. In the lower graph, the locally strongest triplet (singlet) transitions are connected by blue (red) lines, representing transitions to triplet (singlet) dominated states.

We first discuss the one-photon process corresponding to the Stokes pulse, which is shown in the upper graph of Fig. \ref{x-a-leg}. The rovibronic ground state is  localized at the bottom of the $X^{1}\Sigma^{+}$ potential, which has its minimum at $R_{eq,X} = 3.499 \text{\AA}$. In the region in which this singlet wavefunction has significant amplitude, the singlet transition dipole moments do not vary by more than $10 \%$, being 9.7\,Debye $(A^{1}\Sigma^+)$  and 7.5\,Debye (B$^{1}\Pi)$ at the equilibrium distances~\cite{aymar}. The ground state wavefunction then acts as a Gaussian filter between 3 and 4 $\text{\AA}$ for the singlet part of the intermediate states. The first eight eigenstates around -7000\,cm$^{-1}$ in Fig. \ref{x-a-leg}  have dominant triplet character, resulting in modest matrix elements. As soon as the bottom of the A$^{1}\Sigma^{+}$ PEC is reached, the matrix element rises steadily due to the  increasing Franck-Condon factors. When the inner turning point of the A$^{1}\Sigma^{+}$ state approaches $R_{eq,X}$, the matrix element maximizes at 2.92 Debye for the eigenstates around $E \approx -5260\,\mathrm{cm}^{-1}$ and decreases rapidly thereafter. Due to the large electronic dipole moment of the X$^1\Sigma^{+}$ to A$^{1}\Sigma^{+}$ molecular transition, its peak value gives the global maximum moment achievable for the Stokes pulse. A similar behavior is observed around $E_{0}$, when B$^{1}\Pi$ contributions enter into the intermediate states and the matrix element peaks with an absolute value of 2.52\,Debye at $E \approx -900\,\mathrm{cm}^{-1}.$ In addition to the two singlet dominant structures, one recognizes the window (blue color in Fig. \ref{x-a-leg}) between the potential minima of the c$^{3}\Sigma^{+}$ and the B$^{1}\Pi$ state. In this region, a significant admixture of B$^{1}\Pi$ character to the dominantly  c$^{3}\Sigma^{+}$ molecular states leads to significant transition dipole moment from the purely singlet X$^1\Sigma^{+}(v=0,J=0)$ state to the  c$^{3}\Sigma^{+}$  dominated intermediate states. Note that transition matrix elements to these triplet dominant states largely exceed matrix elements to close lying singlet dominant states. 

The lower graph of Fig.\,\ref{x-a-leg} shows the matrix elements needed for the Pump pulse for the singlet and triplet domain, respectively. Due to the large extension of the loosely bound Feshbach molecules over considerable internuclear distances, the wavefunction overlap will in general tend to be the larger the closer the intermediate level gets to the dissociation limit. This is strongly noticeable for the singlet fraction, which shows a sharp rise in dipole matrix element around $E\approx-4500 \mathrm{cm}^{-1}$ due to largely increasing Franck-Condon factors. At lower energies, the singlet dipole matrix elements are below $10^{-6}$ Debye and the singlet component can be considered negligible. At $E \approx -2530\,\mathrm{cm}^{-1}$, the triplet matrix element rises by one order magnitude, marking the beginning of the c$^{3}\Sigma^+$ dominated intermediate states. Matrix elements corresponding to these transitions are about one order of  magnitude larger than the ones which can be assigned to $^{3}\Pi$ transitions, owing to the fact that the $^{3}\Sigma^{+}$ electronic transition dipole moment is considerably larger in the region of interest, which is further enhanced by a higher Franck-Condon overlap at the outer turning points. The largest values appear for levels at and beyond the $P_{\nicefrac{1}{2}}$ asymptote ($E = -57.72$\,cm$^{-1}$). However, the density of states is very large in this area. In addition,  phenomena like predissociation have to be taken into account, opening unfavorable decay channels. We will discard because of the complex unfavorable structure the $^{1}\Pi-{}^3\Sigma^{+}$ crossing resonances, which appear roughly 10\,cm$^{-1}$ below the $P_{\nicefrac{1}{2}}$ asymptote.

\begin{figure}
\centering
\includegraphics*[width=0.95\columnwidth]{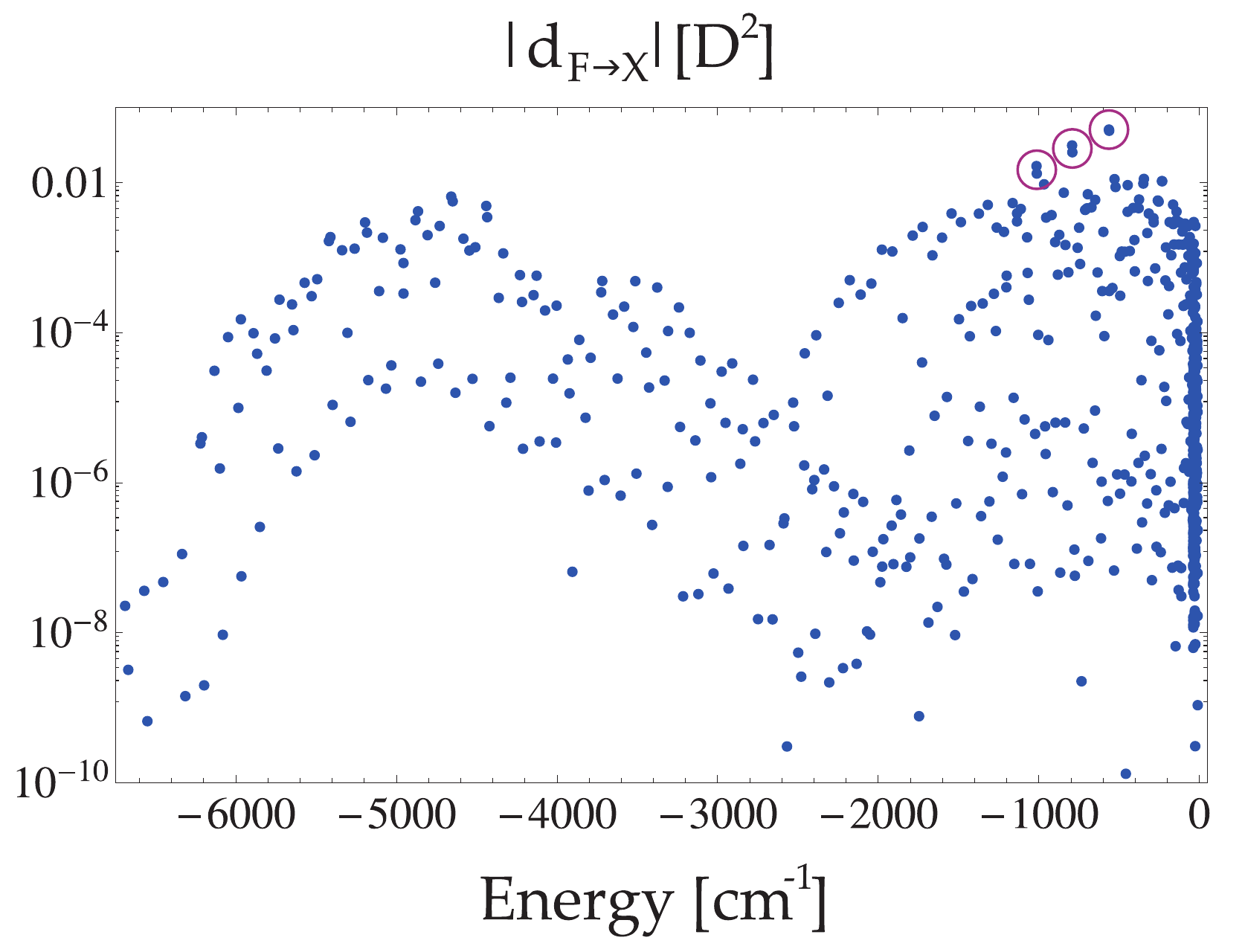}
\caption{(Color online) Absolute value of the two-photon dipole matrix element. The circles mark double dots indicating
degeneracy induced resonances.}
\label{ramanleg}
\end{figure}

\begin{figure}
\centering
\includegraphics*[width=1.0\columnwidth]{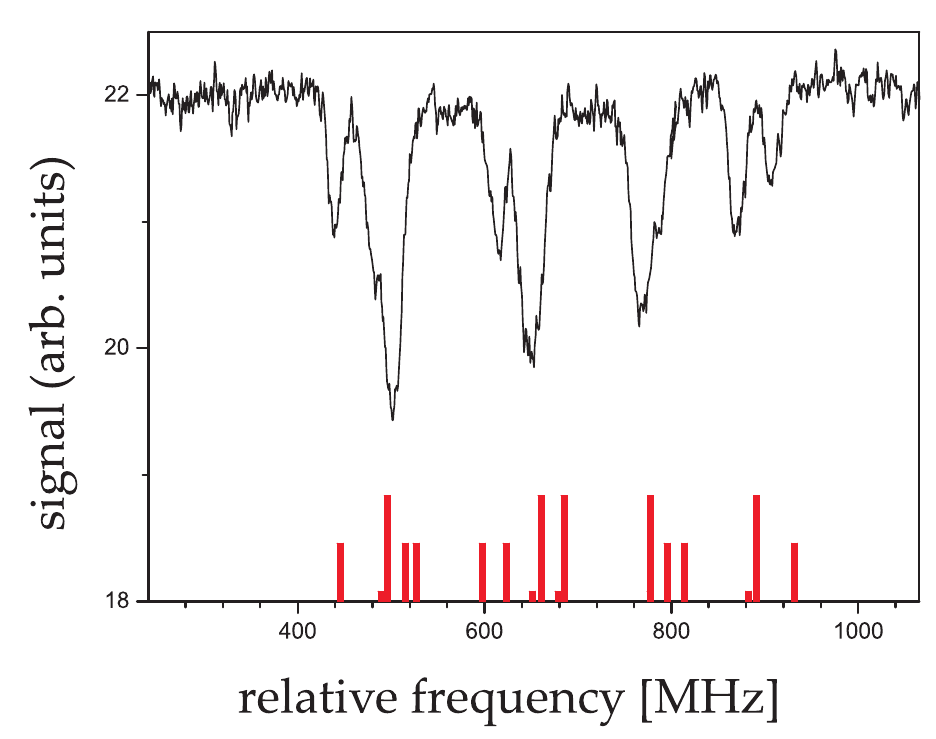}
\caption{(Color online) Part of the hyperfine spectrum of the v=5,N=6 level of the a$^{3}\Sigma^{+}$ state
by STIRAP transfer ($X^{1}\Sigma^{+},v=0 \rightarrow $ intermediate state $ \rightarrow a^{3}\Sigma^{+}$) on a molecular beam of NaK.}
\label{BeamSpec}
\end{figure}
In Figure \ref{ramanleg}, we show for all intermediate states the two-photon transition dipole matrix element $d_{(\mathrm{F}\rightarrow\mathrm{X})}$ obtained from the matrix elements shown in Fig. \ref{x-a-leg} adding singlet and triplet contributions. Degeneracy induced resonances can be recognized due to the appearance of double dots (examples are encircled in figure \ref{ramanleg}). The largest transition dipole matrix elements are obtained at one of these resonances situated in the $^{3}\Sigma^{+}-{}^1\Pi$ regime, where two-photon matrix elements of $0.052 D^{2}$ can be reached at $E \approx -560\,\mathrm{cm}^{-1}$ making these resonances highly interesting for two-photon transfer schemes. However, the exact positions and properties of degeneracy induced resonances critically depend on small corrections of excited state molecular potentials and can therefore not be predicted theoretically. Making use of these resonances demands accurate spectroscopic knowledge of the resonance positions as it can only be obtained by experiments. We have spectroscopically identified one of such resonances in our molecular beam experiment and subsequently used the resonance for a coherent STIRAP transfer from the X$^1\Sigma^{+}$ ground state to high lying vibrational states of the a$^3\Sigma^{+}$ molecular potential. The realized scheme is the reversed process compared to that drawn in figure \ref{fig0}. The pump laser was fixed to the transition from X$^1\Sigma^{+}$(v=0, J=6) to a selected resonantly mixed B$^1\Pi$ $\sim$ c$^3\Sigma^{+}$ level. While the Stokes laser was tuned across the c$^3\Sigma^{+}$ - a$^3\Sigma^{+}$ transition the STIRAP transfer was observed as a reduction of fluorescence out of the upper level. As an example we show in figure \ref{BeamSpec} part of the observed dip structure due to the hyperfine structure of the v=5, J=6 level of the a$^3\Sigma^+$ state. The vertical bars below indicate the expected hyperfine pattern employing atomic hyperfine parameters, where the lengths of the bars only indicate the sum of unresolved levels. Details on this experiment will be given in a forthcoming paper. Extending this experiment towards the Feshbach molecular states and reversing the process and states, as originally given in figure \ref{fig0}, will result in one of the resonance enhanced pathways for the creation of ultracold ground state NaK molecules from Feshbach molecules.

Further possible pathways can be identified by our theoretical analysis which provides a robust description of two-photon dipole matrix elements away from the resonances. Intermediate c$^{3}\Sigma^{+}$ levels perturbed by a small admixture of singlet character from the B$^1\Pi$ molecular potential lead to two-photon coupling matrix elements from the Feshbach state to the X$^1\Sigma^+(v=0,J=0)$ state of $10^{-5}-10^{-2} D^{2}$, coming close to the values obtained for degenerate resonances. The favorable cases are found in the energy region given by blue dots in the middle of Fig. \ref{x-a-leg}. It is further noted that in this area, the one-photon dipole moments for the two involved transitions can be of similar magnitude. In general, a STIRAP sequence benefits from similar peak Rabi frequencies of Pump and Stokes pulse in terms of robustness. We find such states e.g. at $E \approx -1542\,\mathrm{cm}^{-1}$, when the singlet (triplet) dipole matrix elements read 0.057 (0.069) Debye. For the ground state case, this exceeds the 0.046 Debye reported in a similar analysis for the KRb case \cite{kotochigova}. We therefore conclude that such states will be highly promising candidates for the two-photon Raman process.

\section{Conclusion}
In summary, we have presented a multichannel analysis of electronic ground and the excited state manifold converging to the K($4p$)+Na($3s$) dissociation limit of NaK. Our analysis includes a detailed study of the spin character and the molecular wavefunctions of both the Feshbach molecular state and the electronically excited intermediate states. For the Feshbach molecular wavefunction, we do observe a strong magnetic field dependence which can be used to enhance the amplitude of the Feshbach wavefunction in the FCF relevant region and therefore enhance the transition strength from the Feshbach to the intermediate state. For the intermediate electronically excited state, however, we do not expect a strong magnetic field dependence, as the expected Zeeman energy is far weaker than the spin-orbit interaction and the vibrational spacing. 

Our study allows to identify  possible coherent two-photon transfer pathways from Feshbach molecules to rovibronic ground state molecules in a fairly wide energy range accessible by conventional laser sources. The study is complemented by a first demonstration of STIRAP transfer from the vibrational ground state of the \Xstate molecular potential to the \astate manifold demonstrating singlet-triplet transfer in the NaK system via a resonantly mixed excited state. Our analysis therefore fills a critical gap towards the creation of chemically stable ultracold NaK molecules.

\section{Acknowledgements}
We acknowledge financial support from the Centre for Quantum Engineering and Space-Time Research QUEST and the European Research Council through ERC Starting Grant POLAR. E.T. acknowledges the support from the Minister of Science and Culture of Lower Saxony, Germany, by providing a Niedersachsenprofessur. T.A.S. acknowledges financial support from HALOSTAR, M.G. and T.H. from the Research Training Group 1729 of the DFG.

\bibliography{NaKv07}

\end{document}